\documentstyle[12pt,fleqn,epsf,epsfig]{article}
\begin{document}
\title{ Life as Complex Systems: Viewpoint from
Intra-Inter Dynamics}
\author{Kunihiko Kaneko\\
{\small \sl Department of Pure and Applied Sciences,
College of Arts and Sciences,}\\
{\small \sl University of Tokyo,}\\
{\small \sl Komaba, Meguro-ku, Tokyo 153, Japan}\\
}

\date{}

\maketitle

\begin{abstract}

Basic problems in complex systems are surveyed in connection with Life.
As a key issue for complex systems, complementarity between
syntax/rule/parts and semantics/behavior/whole is stressed.
To address the issue, a constructive
approach for a biological system is proposed.  As a construction in a computer,
intra-inter dynamics is presented for cell biology, where
the following five general features are drawn from our model
experiments; intrinsic diversification, recursive type formation,
rule generation, formation of internal representation, and
macroscopic robustness.  Significance of the constructed logic
to the biology of existing organisms is also discussed.

\end{abstract}

\section{ Complex Systems}

In the recent trend, complex systems have often been studied as a problem of 
self-organization or adaptation.  It is typically seen in the 
term `complex adaptive systems', while
the long-term studies for self-organization have been pursued in 
Brussels group as `dissipative structures' \cite{NicolisPri} 
and at the Stuttgart group
as `synergetics' \cite{Haken}.  As will be discussed 
in the present paper, `complex system' studies
should be distinguished from such self-organization or adaptation\footnotemark.

\footnotetext{ Complex system studies in Japan, 
started around mid 80's, are aimed at the understanding of life
by going beyond `chaos' or 'self-organization'\cite{KKTSUDA1,KKTSUDA2}.}

To discuss the problem of `complexity', one should carefully distinguish
`complex' from `complicated'.
The latter is a system composed of a variety of elements, which
requires hard efforts to disintegrate it into parts, but in principle
it is possible.  On the other hand, in `complex systems',
one has to face the circular situation in which
the parts are understood only through  the whole, although
the whole, of course,  consists of parts.  We call such
a structure as `complementarity' \cite{Bohr} between the whole and parts.

Indeed such complementarity is extended to
that between a group of {\sl symbol, rule, syntax} and  that of
{\sl image, behavior, semantics}. (see Table I) \cite{KK-Ike}.
The former group is given by discrete representation, while the latter uses
continuum.  For most `so-called' complex systems,  people still focus on the
studies from the former to the latter.  For example
chaos and cellular automata studies have shown how 'complex behavior' 
emerges from a simple rule, while the rule formation
from dynamic behavior is not well discussed.  In complex system studies,
we also have to study how the former set (i.e., syntax etc.)  
emerges from the latter (semantics etc.).  

There are very few studies towards this direction, although 
in chaos studies, there are some germs for it.  In Crutchfield's 
`$\epsilon$-machine', he tries to construct an automaton rule from 
data\cite{Crutch}.  In several high-dimensional dynamical systems, chaotic 
itinerancy \cite{GCM,CI1,CI2} 
over ordered states have been found, where the transition from one ordered state
to another is not random, but is constrained with some rule generated from
lower-level dynamics.
Indeed frequent appearance of chaotic itinerancy in biologically 
oriented problems gives us a hope that the formation of a syntactic rule
from complex behavior may be associated with it.

\vspace{.2in}
Table I

\vspace{.1in}
\begin{tabular}{|c||c|c|c|} \hline
parts & whole \\ \hline
efficiency of parts & macroscopic robustness \\ \hline
digital logic & analogue pattern dynamics \\ \hline
syntax & semantics \\ \hline
symbol & image \\ \hline
rule of gene switch & body plan (pattern) \\ \hline
\end{tabular}

\vspace{.2in}

Let us take an example in molecular biology.  It is now generally
believed that genes switch on and off, according to signals.
The mechanism of on-off switching has been studied in detail \cite{Cell}.
As a combination of such local switching process the body
plan is described. Still the question remains if this
mechanism is complete as explanation.
In spite of such `mechanical' explanation there, biological processes
in cells occur in a thermodynamic fluctuation, and
explanation on the robustness against fluctuations is required.
Even if some errors such as somatic mutations occur, the
total body often remains robust against such errors.
If one assumes that cells change their fates passively
under externally given signals,
instability in the developmental process remains, since
tiny perturbations can change their cells' fates
according to the switching mechanism with a given threshold.
Then we need another mechanism to modify local rules so that global 
developmental process is robust\footnotemark.  One might describe such mechanism
in terms of molecules and add a new complicated rule.
This process of searching for and adding a new local process
can continue for ever so that the 
molecular biology looks complete.  Still one may wonder if this
is the ultimate explanation.  

\footnotetext{Such mechanism to correct error, however, is under
molecular fluctuation again.  Hence a logic leading to global robustness
is postulated \cite{KK-TY3}.}

One has to note that organisms are not
designed by somebody, but have evolved spontaneously.  
In man-made machines, 
parts, chosen to have efficient function, are combined to work as 
designed so that no serious damage is given by the interaction among
the parts.  The principle of organisms can be different from such machines.
The function of each part changes in relationship with
other parts.  Rules that each part obeys are neither designed nor given in 
advance.  They are context-dependent and are variable according to the 
information on the whole.  We need some logic why such rules appear 
generically for a class of biological systems.

\section{Our approach}

\subsection{constructive biology}

If a biological system were just a machine with complicated
combination of parts, our possible study would be just to describe the details 
of elementary processes in the present organisms, by abandoning the attempt
to find any universal logic therein.  On the contrary,
we believe that organisms are complex 
systems, and that there exists
universal logic leading to their common nature.

Note that approaches for complex and complicated systems should be
distinguished.  Since the latter are essentially understood as
a combination of simple processes, what should be done here is
to search for minimal sets of local processes that can fit
real data.  On the other hand, for complex systems
( in the sense of complementarity in \S 1), such approach
is not effective.  One has to search for a general logic
why such complex system is of necessity and universal.

Here it should be noted that the logic that organisms
necessarily should obey is not completely revealed as long as we 
study only the present organisms 
( or slight modification to them by mutations).
The approach that should be taken will be 'constructive' 
\cite{KKTSUDA1,KKTSUDA2} in nature.
We combine several basic processes, and construct a class of models,
to find universal logic underlying therein. With this logic,
biological systems are classified into some universality classes.
The present organisms, then, are understood as
one representative for a universal class, to which the 
life-as-it-could-be also belongs. 

Our `constructive biology' consists of the following steps.
(i) construct a model by combining procedures;
(ii) clarify universal class of phenomena through the constructed model(s);
(iii) reveal the universal logic underlying the class of phenomena
and extract logic that the life process should obey;
(iv) provide a new look at data on the present organisms
from our discovered logic.  There are
two possible ways, to perform these steps.

The first one is the construction of artificial world in a computer 
by combining well-defined simple procedures, and extraction of a general 
logic therein \cite{KK-TY1,KK-TY2,KK-TY3,KK-Keio}.
It should be noted that we do not intend to imitate the 
life.  Instead we search for a universal class of behaviors, and reconstruct
the present life as an example of such class.
Strategy of such construction will be discussed in later subsections.

The second approach is experimental.  In this case again, one constructs a  
possible biology world in laboratory, by combining several procedures.
For example, this experimental constructive biology has been
pursued by Yomo and his collaborators \cite{Yomo} at the
biochemical reaction level,
the organism level, and at the level of ensembles of cells.

In this respect, our constructive biology has something common
with the so-called Artificial Life\cite{AL}.  Still, there are two differences.
First, we are strongly motivated to search for universal logic
underlying the constructed world.
The second difference is our emphasis on the complementarity structure.
The so-called artificial life focuses on the  emergence of complex behavior 
from syntactic structure.
In our approach for complex systems, the other direction, i.e.,
the emergence of syntactic rule from complex behavior 
is pursued as mentioned.  
It should also be emphasized that we intend to 
reconstruct the present organisms as a class of constructed world,
and propose novel viewpoints to the `real' biology.

\subsection{dynamical systems}

Since we aim at constructing a scenario from the image/pattern to
symbol/rule, we need some system to describe the change in analogue
pattern.   For this we adopt a dynamical system approach
(in its broadest sense).  A dynamical system consists of
time, a set of states, an evolution rule, an initial condition of the states,
and boundary conditions.  The state is represented by a set of $k$ variables,
the ``degrees of freedom".  Thus the state at an instant is
represented by a point in the $k$-dimensional space called phase space.

It is generally assumed  
in a dynamical system that a set of state variables, an evolution
rule, and initial and boundary conditions of the states are given independently
of each other.  The states evolve
according to the rule, but the set of states itself 
( e.g., the number of variables) is fixed.  The states cannot change the
evolution rule itself. Choice of
initial and boundary conditions are independent of the
state and of the rule.

In biological problems, however, such separation between all these
elements of the model may not be valid from the beginning \cite{CML-cell}.
The evolution rule itself is formed and changes in connection with
the temporal evolution of the states.  A simple example
is the change of  the number of variables itself with time:
In the development of a cell society,
when each cell state is represented by
a set of variables, an extra set is demanded by cell division,
unless the two cells remain identical.
This change in degrees of freedom is dependent on the
state, since the cell divides or dies according to its state.
Hence we have to study `open dynamical systems', in the sense
that the phase space dimension changes according to the state.
Some mechanism for the 'dynamics of dynamics' has to be introduced,
to incorporate with the interference between the evolution rule and states.

Among the open dynamical systems, we have coined the term
``open chaos" to discuss the change of the dimension associated with
chaotic instability \cite{KK-AL}.  Indeed, in a model of cell differentiation to be
discussed, difference between cells is amplified by orbital instability.
If two cells remained identical, we would not need another set of variables.
Thus the increase in the number of variables is tightly connected with the
orbital instability, as in chaos.
It should also be noted that in the developmental process, in general, 
initial and boundary conditions of states
are chosen so that reproduction continues, from their mother's states.

\subsection{intra-inter dynamics} 

What type of dynamical systems  should one construct to capture the 
basic features outlined in \S 1?  
First, we assume there is a unit separated from the outside.
This is not trivial, since in nonlinear systems, tiny change in one element
may be amplified to other elements, and they are not
separated each other.  As long as we take
the importance of interactions for granted, the formation of a unit itself
is nontrivial.  In other words,
the separation of the unit from surroundings cannot be
complete in a nonlinear system, and holds only approximately.  It is made 
possible only by forming
an interface between the outside and internal structure, by which
some information on external states are embedded inside.
In the term of cell biology, this corresponds to the origin of cell,
an important step to life, which is made
possible by the existence of a membrane. It makes a boundary between
the inside and the outside, but in a rather flexible way.

A biological unit thus formed must have  always
internal structure.  Furthermore, these units strongly interact with 
each other, due to incomplete separation.
Hence we need a model consisting of
the interplay between inter-unit and intra-unit dynamics \cite{CML,GCM}.
For example, complex chemical reaction dynamics
in each unit (cell) is affected by the interaction among cells,
which provides an interesting example of
``intra-inter dynamics".

As a specific example of the scheme of intra-inter dynamics, we mainly
discuss the development process of a cell society accompanied by
cell differentiation \cite{Turing}.
Here the intra-inter dynamics consists of several biochemical reaction 
processes, while the interaction is inter-cellular through diffusion of 
chemicals, other signal transmission, and so forth.  The change of dynamics
itself is brought about by cell division and death, depending on
the cellular state, by which the degree of freedom varies.

\subsection{our model}

Now the remaining questions are
choice of
(a) the internal variables and their dynamics,
(b) interaction, and
(c) a rule to change the degrees of freedom (e.g., cell division).
There is a variety of possibilities of models according to
these choices, and we have studied several of them.

Although the details of our model are given in the previous publications
\cite{KK-TY1,KK-TY2,KK-TY3,CF-KK,KK-Rel,CML-cell}, it may be helpful to
explain the basic structure of our modeling briefly.

As a set of variables, we take concentration of a set of chemicals.
For the internal dynamics, auto-catalytic chemical reaction among these
chemicals are chosen. This auto-catalytic
reaction mechanism is necessary to produce some chemicals in a
cell ( see also \cite{Eigen-Schuster}).  Note that such auto-catalytic
reaction dynamics often leads to nonlinear oscillation in chemicals.

As for the interaction, diffusion 
of chemicals between a cell and its surroundings is considered.
The surrounding medium is assumed 
to be spatially homogeneous in most simulations,
to avoid the complication by spatial pattern.  In other words,
all cells couple to all others.  For some models,
we assume a nutrition chemical, and its active transport into 
a cell, whose rate depends on the concentration of some chemicals
within the cell.

The condition of cell division is written as an integral form.
In a class of model, we assume that some products are accumulated through the
chemical reaction.  When the
accumulated product concentration goes beyond some threshold,
the cell divides into two.
In another class of model, cell volume is computed according to the
chemicals within, and a cell divides into two when the volume becomes
the twice of the original.

Of course, there can be a variety of choices on the
chemical reaction network.  Indeed, the observed results
do not depend on the details of the choice, as long as the
network allows for growth in cell numbers. 
Note that we have not constructed our network based on some
data on biochemical network.
Rather, we try to demonstrate that important features in a biological system 
are a consequence of a system with internal dynamics, 
interaction, and reproduction.  From the study we have found a
universal logic underlying this class of models.
Later we will survey the logic in connection with cell biology,
although the scenario itself is believed to hold generally
in a biological system.

\subsection{remarks}

Note that our approach for development is intended to make a connection
from `pattern' to `rule' in Table I in \S 1.  As for the study of the 
development at the pattern level (right hand side of the Table), there is 
Turing's pioneering study
while the study from a digital rule (from the left hand side of the Table)
is pioneered by
Kauffman\cite{Kauff} and Lindenmayer \cite{Linden}.
Before discussing our scenario and logic of the development
of a cell society, it may be relevant to comment on the  previous
theoretical studies on the differentiation and development.

Turing proposed a pattern-formation mechanism according to the
instability in a reaction-diffusion system.  
In this sense, Turing's
study aims at understanding the morphogenesis at the right side
of Table I.  On the other hand, we aim at reaching the rule formation
starting from the right side.  Note also that inclusion of threshold mechanism
on the diffusion mechanism \cite{Wolpert}, which may be regarded as
a variant of Turing's idea, is not sufficient to form a process from
the right-hand side to the left in Table I. See also
\cite{Newman} for a connection from Turing pattern to a genetic rule.

In contrast with these `cell-to-cell interaction'-based study,
differentiated cell types are attributed to different attractors in
an intra-cellular dynamics, in some other studies.  For example, in
Kauffman's study \cite{Kauff},  coexistence of many attractors in
a Boolean network corresponds to a different cell type.
Attraction to different states is
found in a coupled system of Boolean-network-type
differential equations \cite{Glass-Kauff},
while a CML corresponding to  such Boolean networks
is studied by Bignone\cite{Bignone} ( see also \cite{reinitz}).
As will be seen later, our cell type is not represented by
an attractor, but by a state stabilized through cell-to-cell interaction.

In our approach we assume temporal oscillations of
intra-cellular chemical dynamics.  The importanceof oscillations
was pioneered by Goodwin\cite{Goodwin}, and
the existence of oscillatory dynamics for
cell division processes has been discussed both experimentally and 
theoretically.  In connection with the development process,
Goodwin and Cohen \cite{Goodwin-Cohen}
proposed a mechanism of positional information,
through the phase difference of intra-cellular oscillations. 

As will be seen, our scenario is distinguished from previous studies,
by the importance of orbital instability, 
stability at an ensemble level,
change of cell numbers in conjunction with dynamics,
formation of recursive types and differentiation rules.

\section{Logic for Diversification, Memory, and Robustness}

From several simulations of the models
starting from a single cell initial condition, Yomo and the author
proposed the ``isologous diversification theory", as
a general mechanism of spontaneous differentiation of replicating
biological units \cite{KK-TY1,KK-TY2,KK-TY3}.   From the
differentiation process of the models, we extract the following basic
features; (i) intrinsic diversification,
(ii) type formation, (iii) recursivity (iv) rule formation 
and (v) macroscopic robustness.  Here we will discuss these features
from the viewpoints of model results, the logic that supports them,
and their significance to biology.
(For other related issues see \cite{KK-Keio}).

{\bf (1)  Intrinsic Diversification}

[ Observation from our Model]

Up to some number of cells, all cells are identical as to the chemical
concentration, which oscillate coherently.
Accordingly, the cells divide almost simultaneously, and the number of cells
is a power of two.
When the number of cells exceeds some threshold, they lose identical and
coherent dynamics.  Cells separate into several groups whose phases of 
oscillations are close.
At this first stage, only the phases of oscillations are different by cells.
Cells are not differentiated yet, since 
the temporal averages of chemicals, measured over periods
of oscillations, are almost identical.  After further divisions of cells,
chemical compositions start to be different by cells and the differences 
remain.  At this second stage the chemical concentrations differ by cells,
even after taking the temporal average
over periods. Thus the behavior of states  is differentiated.
Here the orbits of chemical dynamics lie
in a different region by groups within the phase space.

[Logic]

The change of phases at the first stage is due to clustering,
studied in coupled  nonlinear oscillators\cite{GCM}. Clustering 
generally appears
in a coupled system as long as there is both orbital instability to
amplify small differences and a tendency to keep the synchronization.
The differentiation at the second stage, with the separation  
in the phase space, is expected if the instability is
related also to the amplitude of oscillations, and some internal
degrees of freedom exist to support the difference in
the phase space position.  We should note that
the diversification is a general consequence of coupled nonlinear systems.

[Biology]

Note that the cell has an intrinsic trend for diversification.
Indeed, it is hard to imagine that a complete replication machinery
existed at the early stage of life.  Furthermore, all cells
are different in our body, strictly speaking.  Cells tend to
diversify at the early stage of a developmental process.

The diversification is relevant to the effective use of finite resources.
As has been discussed \cite{KK-Rel,KK-TY1}, phase
clustering provides
time sharing for resources: Cells can get a chemical resource successively in
order, according to the difference in phase of oscillations.  
The differentiation of chemical compositions supports the 
differentiation of roles.
Such differentiation is seen at a rather primitive stage
of multi-cellular organisms, such as the Volvox \cite{volvox}.

{\bf (2)  Recursive Type Formation}

[Observation from our Model]

After fixed differentiation, the chemical composition
of each group is inherited by their daughter cells.
The determination of a cell has occurred
at this stage.  The cell state represented by average chemical composition
remains identical by division,
and thus the  cells keep the ``recursivity" by divisions.
The chemical characters are ``inherited"
just through the initial conditions of chemical concentrations after the
division, although we have not explicitly imposed
any specific mechanisms to keep the type.
In other words, a kind of memory is formed.

[Logic]

The cellular memory at the above stage is formed as a result of the
selection of initial conditions for a cellular state.
In our model all cells are coupled  and form
a single dynamical system. Each cellular state does not constitute
an independent dynamical system.  Still, it is effective to define an
initial condition of each cell state forming a `partial dynamical system'
out of the whole.  The recursivity is attained through the selection of initial
conditions of such a partial system, so that it is rather
robust against change of interaction terms also.
This mechanism for recursivity is possible
only in an open dynamical systems that allows for change of degrees
of freedom.

[Biology]

In spite of the first diversification trend, biological units (e.g., cells)
tend to be classified into several types, whose number is much smaller
than the number of units.  With growth in numbers,
new-born units (cells)
themselves are not identical with their ancestors, as mentioned,
but after some generations, they tend to belong to the same type.
Although cells of the same type
are not completely identical with each other, the difference among them is
much smaller than that between cells of different types.
In the developmental course, cells are initially undifferentiated, and change
their states with time.  Later,
fixed cell types appear, called determination \cite{Cell}.

Indeed such a tendency to form discrete types is not limited to
cells.  Organisms seem to be classified into types,
which may often be called ``species".  Currently ``species" are
defined through the sterility of hybrids, leading to sexual 
isolation. However, we should note that such species-like types also exist
in asexual organisms, and even in uni-cellular organism.
Hence we need some logic that supports the type formation beyond
sexual isolation.

{\bf (3) Rule Generation}

[Observation from our Model]

As the cell number increases, further differentiation proceeds.
Each group of cells further differentiates
into two or more subgroups.   Some cell-types continue to switch to
others besides replication of the same type.  Here the switching
follows some rule, by which hierarchical differentiation proceeds.

For example, six types ("S","A","B","A1","A2","A3") are formed
successively in a model \cite{CF-KK}.  The orbits of
each type are clearly separated from each other in the phase space of
chemical concentrations.
Chronologically speaking, the type "S" first appears and
replicates.  When the number goes beyond some threshold, switching
from S to A or to B starts to occur at some rate.  With further
divisions, switching from A to three types A1,A2, and A3 occurs.
The chemical compositions of these three types are not much different
from A, as compared with the differences to B or S.
The differentiation rule here is found to obey the rule shown in Fig.1.
($S\rightarrow S,A,B$ $|$ $A\rightarrow A,A1,A2,A3$ $|$ $B\rightarrow B$ $|$
$A1\rightarrow A1$ $|$ $A2\rightarrow A2$$|$ and  $A3 \rightarrow A3$),
in the normal course of differentiation starting from a single cell.
A hierarchical rule of differentiation is thus
generated.  Formation of this kind of rule
is generally observed in a class of chemical networks.

[Logic]

This differentiation rule that the cells obey is given in their internal 
state and interaction.  In the above case, the orbits corresponding to the 
'S' state has paths to the states A and B, in the presence of
sufficient number of other cells.
The rule is given in the orbits in the internal dynamics, which is
modulated by the interaction.  This rule is a higher-level one
than the original dynamical systems (i.e., chemical reaction rule).
Formation of such higher-level rules reminds us of 
chaotic itinerancy, where switches among several ordered states
(with lower degrees of freedom) emerge out of high-dimensional chaotic
dynamics\cite{GCM,CI1,CI2}.

[Biology]

Stem cells either replicate or
differentiate into different cell type(s).  This differentiation
rule is often hierarchical, as in Fig.1 \cite{Cell,Ogawa}.  Such rule is 
expected to be not solely determined by internal
cell states.  Otherwise, it is hard to explain why the development
process is robust.  For example, when the number of some terminal cells is
decreased, there should be some regulation to increase the rate of the
differentiation from the stem cell to the terminal cells.
This suggests the importance of interaction, as in our
model results.

{\bf (4) Internal Representation}

[Observation and Logic]

As for cell types,  one might think that this selection is 
nothing but a choice of basin of attractions
for a multiple attractor system.  If the interaction were neglected,
this would be basically correct.  In our case, this is not true.
Indeed most dynamical states of each cell type do not exist as an
attractor but are stabilized through interaction.
The observed memory lies not solely in the internal states but
also in the interactions among the units.

To see the intra-inter nature of this memory explicitly,
one effective method is a transplantation experiment.
Numerically, transplantation experiments are carried out
by choosing determined cells (obtained from the normal differentiation process)
and putting them among a different set of surrounding cells,
to make a cell society that could not appear through the normal course
of development.

When a determined cell is transplanted to another cell society,
the offspring of the cell remain the same type,
unless the cell-type distribution of the society is strongly biased
( e.g., the ensemble consisting only 
of the same type of the cell as transplanted).
The cell memory is preserved mainly in each cell,
but suitable cellular interactions are also necessary to keep it.
The achieved recursivity is understood as the choice of
internal dynamics through cellular interactions.

Depending on the distribution of other cell types,
the orbit of internal cell state is deformed, and leads to a switch
to other cell types if the deformation is large.  Such modulation is
possible, due to our `dual' memory system:  A cell's state is mainly 
characterized by its discrete types, while there remains analogue
modulation even among the same cell types, which reflects on the
global information on cell distributions. 
Internal representation of other cells is formed with the
modulation of each cell.

Such continuous deformation (modulation) is also seen in the model for a stem 
cell, where cell types switch to further subgroups with some rate \cite{CF-KK}.
The rate of the differentiation or the replication
( e.g., the rate to select an arrow among $S\rightarrow S,A,B$)
depends on the cell-type distribution.  For example
when the number of ``A" type cells is reduced, the orbit of an ``S-"type cell
is shifted towards the orbits of ``A", with which the
rate of switch to ``A" is enhanced.  In this case,  again,
the information on cell number distribution is represented by
the internal dynamics of ``S"type cells.

{\bf (5) Macroscopic Robustness}

[Observation from our Model]

Since our differentiation mechanism includes
instability in internal dynamics triggered by cell-to-cell interactions,
one might then suspect that such a developmental process
would not be robust against perturbations.  This is not the case.
With the above modulation mechanism, 
a  kind of robustness at  the ensemble level is generated.  
The switch of types is regulated so that the distribution of
diverse cell types is restored to the original one.
The global stability of the whole system is thus obtained by
spontaneous regulation of the rates of the differentiation.

[Logic]

In the present case, macroscopic
stability is sustained by the change of the rate of differentiation.
But why is the regulation oriented to keep the stability,
instead of the other direction?
As an example consider the branch in Fig.1. As mentioned,
the differentiation from $S$ to $A$ is enhanced when the type A cell is
removed.  If, the regulation worked in the other way, 
( i.e., the decrease of
the rate $S\rightarrow A$ by the removal of A), then the type-A cell would 
hardly exist from the beginning.
Consider the developmental process:
In the initial stage of the appearance of type-A cells, their numbers
are small.  If the regulation worked to decrease the rate to type-A in the
environment with fewer type-A cells, then the number of type-A cells
would be decreased to zero.  If this were the case, the type-A cell would not appear
from the beginning.
In other words, only the cell types that have a regulation mechanism
to stabilize their coexistence
can appear in our dynamical system.  Note that this
logic is possible since, in our theory, types and  rules of switches are not 
given in advance, but appear as a consequence of dynamics.

In general, stability at the ensemble level is
a rather general consequence of coupled dynamical systems \cite{KK-MF,homeo}.

[Biology]

It should be noticed that global robustness has higher priority than local
rule.  For example, by a mutation to triploidy in the newt, the cell becomes
three times large.  In this case, the total cell number is reduced to one third,
and the final body remains not much affected by the mutation.  In other
words, the local rule of cell divisions, (e.g., the number of divisions) is
modified so that the global body pattern remains undamaged \cite{Cell}.

In general, the developmental course is rather stable against
possible somatic mutations, while in the
hemopoietic system \cite{Ogawa}, existence of regulation mechanism to
keep suitable distribution of cell types is expected.
In the hierarchical structure represented in Fig.1, the cells at an upper
node behave like stem cells, and regulate the number of cells at a lower
node.  This type of regulation system is expected in real
multicellular organisms.

It should be stressed that our dynamical differentiation
process is always accompanied by this kind of regulation process, without any
sophisticated programs implemented in advance.
This autonomous robustness provides a novel viewpoint to 
the stability of the cell society in multicellular organisms.

\section{Discussion}

At the next step in the constructive biology, we have to
provide a new look at present organisms from our standpoint.
Note that our model experiments show correspondence to
tumor cell formation,  stem cells, relevance of chemicals with low
concentrations, germ-line segregation, and so forth.
However, our goal in this constructive biology is
not limited to seeking for such correspondence.  Rather,
we aim at recapturing existing organisms from our
constructed classes of intra-inter dynamics.
So far, we have found the following classes for each unit behavior;

\begin{itemize}

\item
'Undifferentiated cell class'; Cells are not recursive, and their
states change with time, through several switching-type behaviors. 

\item
 'Stem cell class'; Cells either reproduce the same type or switch to
different types with some rate.

\item
 'Germ-line cell class'; After division one of the cells keeps the 
recursivity and high activity, while the other loses activity and 
division is suppressed.

\item
 'Determined cell class';   Cell types are kept by the division.
Note, however, this determination is relative.  When a cell is
determined in the course of development, it can be de-differentiated
and switched to other types by a `radical' change of interaction.

\item
`Tumor-type cell class'; These cells destroy the cooperativity attained in 
the cell society, and grow in numbers in a selfish way.    In 
simulations with a larger diffusion coupling, this type of cells is observed, 
which is an extreme limit of a differentiated cell with specialization
in chemical compositions.  Its chemical configuration loses
diversity, and the ongoing chemical pathway there is simpler than other cell
types.  These cells replicate faster.   Since the emergence of this cell type
destroys the order allowing for diversity in cells, we have called it
tumor cell\cite{KK-TY2}.
Note also that the recursivity by cell division is partially destroyed, 
in the sense that 
the bias in chemical concentrations is preserved to daughter cells, 
but the concentrations
themselves differ by cells\cite{KK-TY2,KK-TY3}.

\end{itemize}

It is hoped that the present cells can be classified according to such 
classes, and general features extracted.  For example, we have proposed
that the tumor cells lose chemical diversity, and also suggested a possible
way to re-differentiate the cells to normal course by restoring 
the diversity\cite{KK-TY2,KK-TY3}.

There are also several possible classes for 
an ensemble of cells.  In this case, behavior is classified according to
the spatiotemporal dynamics of the distribution of cells.  From this viewpoint,
it is again expected that stages of multi-cellular organisms are classified,
starting from the primitive stage by Dictyostellum discoideum,
Volvox, Anabena, $\cdots$, and higher organisms.
It should also be noted that detailed process to form a
cell society is not required in our model study.  This implies that the
origin of a multi-cellular organism is a rather natural consequence
when cells increase in their number and are concentrated to have
strong interaction with each other.

We believe that the current approach is not limited to
cell biology.  One may have noted that our logic for the formation (and
collapse\cite{CML-cell}) of  the cell society may be
extended to human society.  Indeed, temporal switching of roles and class
differentiations are generally seen in socio-dynamics.
An example of the formation of a complex society through such differentiations
is studied in the evolutionary dynamics of iterated games \cite{Akiyama}.

Of course, the complementarity structure between syntax and semantics 
is most important in human cognition processes.   
Internal images are formed so that the emergent rules have
context-dependence. In this sense, our scenario for the
cell society will be relevant to the problem of cognition.

In neural dynamics, we often adopt
the concept of a module, and then try to combine modules to make a total image 
of the world.  It is a typical approach from the left to right in Table I
of \S 1. 
However, the modules are not separated with each other,
and work only as an approximate separation from dynamically connected 
elements.  In this sense, the approach from the right to left in Table I 
of \S 1 has to
be pursued, as has been demonstrated in the present paper. Extension of
our approach to neural dynamics, and to the logic formation from
pattern dynamics will be discussed elsewhere.

\vspace{.2in}

{\sl acknowledgments}

The present paper is based on the studies in collaboration  with T. Yomo and
C. Furusawa. I am grateful to  them for stimulating discussions.  
The work is partially supported by Grant-in-Aids for Scientific
Research from the Ministry of Education, Science, and Culture of Japan.

\pagebreak
\begin{figure}
\noindent
\hspace{-.3in}
\epsfig{file=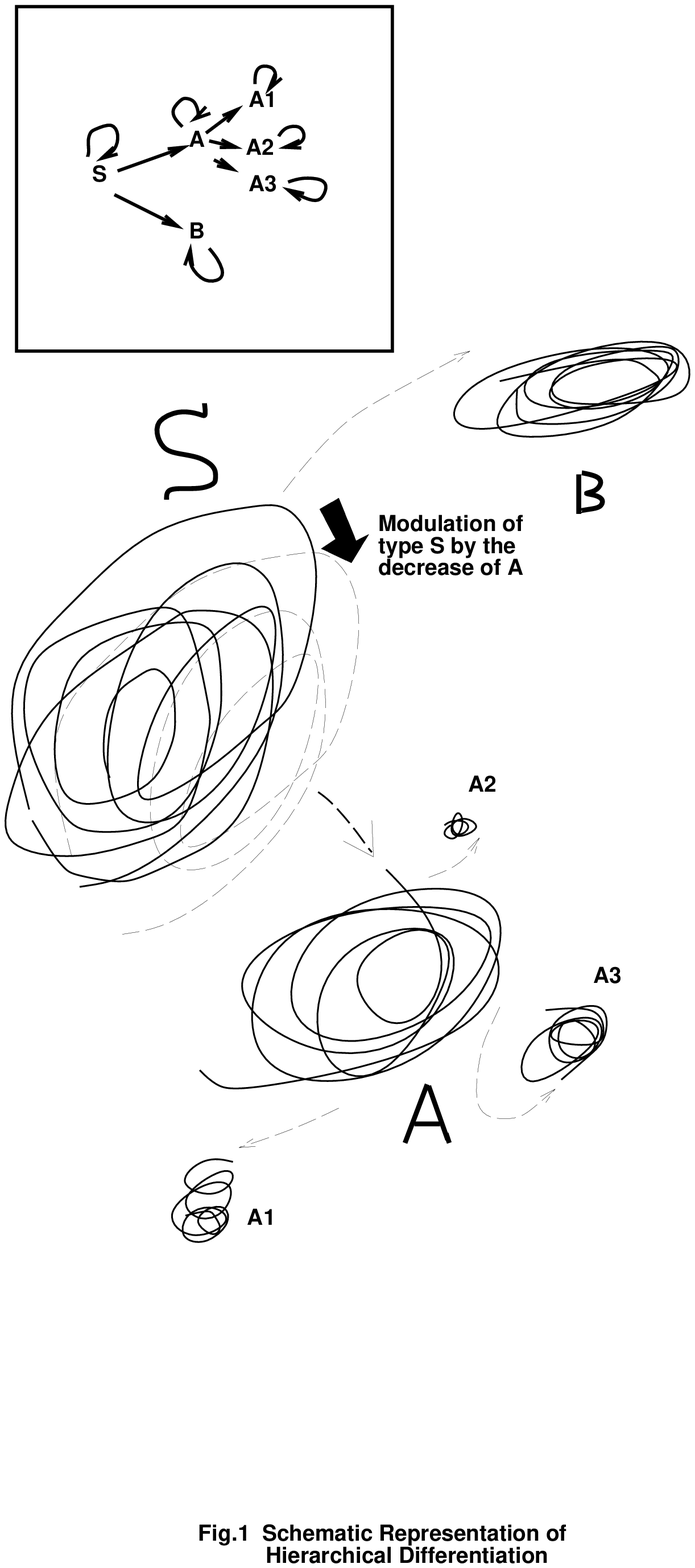,scale=0.8}
\caption{ }
\end{figure}


\addcontentsline{toc}{section}{References}

\begin{thebibliography}{999}

\bibitem{NicolisPri}
G. Nicolis and I. Prigogine, ``Self-organization in Nonequilibrium Systems"
(Wiley, 1977)

\bibitem{Haken}
H. Haken, Synergetics, Springer, 1978

\bibitem{KKTSUDA1}
K. Kaneko and I. Tsuda,
``Constructive Complexity and Artificial Reality: An Introduction",
Physica 75 D (1994), 1-10

\bibitem{KKTSUDA2}
K. Kaneko and I. Tsuda, {\sl Chaos Scenario of Complex Systems},
in Japanese, Asakura pub. 1996 ( English
translation will be published from Springer).

\bibitem{Bohr}
for the general use of this term see N. Bohr, {\sl Atomic Physics and Human Knowledge},
John Wiley, 1958.

\bibitem{KK-Ike}
K. Kaneko and T. Ikegami, ``{\sl Evolutionary Scenario for Complex Systems}"
in Japanese, Asakura pub. 1998.

\bibitem{Crutch}
J.P. Crutchfield, Physica D 75 (1994) 11
``The calculi of emergence: computation, dynamics and induction"


\bibitem{GCM}
K. Kaneko ``Clustering, Coding, Switching, Hierarchical Ordering,
and Control in Network of Chaotic Elements'',
Physica 41 D (1990) 137-172

\bibitem{CI1}
I.Tsuda,"Chaotic itinerancy as a dynamical basis of Hermeneutics in
brain and mind" , World Futures 32(1991)167.

\bibitem{CI2}
K.Ikeda, K.Otsuka, and K.Matsumoto, Prog.Theor. Phys. Suppl. 99(1989)295.
``Maxell-Bloch Turbulence''

\bibitem{Cell}
B. Alberts, D.Bray, J. Lewis, M. Raff, K. Roberts, and J.D. Watson,
The Molecular Biology of the Cell, 1989,1994


\bibitem{CML}
for a prototype of the intra-inter dynamics, see e.g.,
 K.Kaneko ed.,  Theory and applications of coupled map
lattices Wiley (1993). See also \cite{GCM}.


\bibitem{KK-TY1}
K. Kaneko and T. Yomo,
`` Cell Division, Differentiation, and Dynamic
Clustering'',  Physica 75 D (1994), 89-102

\bibitem{KK-TY2}
K. Kaneko and T. Yomo,
``Isologous Diversification: A Theory of Cell Differentiation",
Bull.Math.Biol.  59 (1997) 139-196

\bibitem{KK-TY3}
K. Kaneko and T. Yomo,
``Isologous Diversification for Cell Differentiation", preprint

\bibitem{KK-Keio}
K. Kaneko, ``Diversity, Stability, Recursivity,
and Rule Generation in Biological System:
Intra-inter Dynamics Approach'',
Int J Mod Phys. B., to appear

\bibitem{Yomo}
Ko, P., E., Yomo, T., and
Urabe, I. 1994.  Dynamic clustering of bacterial population.  Physica D
75: 81-88; Wei-zhong Xu, A. Kashiwagi, T. Yomo, and I. Urabe
Res. Popul. Ecol. 38(2) 231-237 1996 ``Fate of a mutant emerging at the
initial stage of Evolution'' Prijambada I. D., Yomo T., Tanaka F.,
Kawama T., Yamamoto K., Hasegawa A., Shima Y., Negoro S., and Urabe I.
1996.  Solubility of artificial proteins with random sequences.
FEBS-Lett. 1996 Mar 11; 382(1-2): 21-5

\bibitem{AL}
C.Langton eds. Artificial Life, II,III (Addison Wesley,1989,1991,1994) 
``Artificial Life: An Overview''(MIT Press 1995)

\bibitem{KK-AL}
K.Kaneko,  `` Chaos as a Source of Complexity and Diversity in Evolution'',
Artificial Life 1, (1994), 163-177G

\bibitem{CML-cell}
K.Kaneko,
``Coupled Maps with Growth and Death: An Approach to Cell Differentiation",
Physica 103 D(1997) 505-527

\bibitem{Eigen-Schuster}
M. Eigen and P. Schuster, {\sl The hypercycle},
Springer Berlin, Heidelberg, N.Y., 1979

\bibitem{Turing}
A.M. Turing, A. M. (1952). The chemical basis of
morphogenesis {\sl Phil. Trans. Roy\ .  Soc.} B, {\bf 237})

\bibitem{Kauff}
S. Kauffman, ``Metabolic
stability and epigenesis in randomly connected nets",
 J. Theo. Biology. 22, (1969) 437

\bibitem{Linden} 
A. Lindenmayer ``Mathematical Models
for Cellular Interaction in Development, I and II"
J. Theor. Biol.  18 (1968) 280.  

\bibitem{Wolpert}
L. Wolpert, J. Theor. Biol. 25 (1969) 1
 Positional Information and the Spatial pattern of
Cellular Differentiation

\bibitem{Newman}
S. A. Newman and W.D. Comper ,
Development 110 (1990) 1-18 Generic physical mechanisms of morphogenesis
and pattern formation

\bibitem{Glass-Kauff}
L. Glass and S.A. Kauffman, J. Theor.  Biol., {\bf 34} (1972) 219.
  "Co-operative Components,Spatial Localization and
  Oscillatory Cellular Dynamics",
  

\bibitem{reinitz}
E. Mjolsness, D.H. Sharp, and J. Reinitz
J. Theor.  Biol., {\bf 152} (1991) 429.

\bibitem{Bignone}
F.A. Bignone, J. Theor.  Biol., {\bf 161} (1993) 231.
 Cells-Gene interactions simulation on a coupled map lattice, 

\bibitem{Goodwin}
B. Goodwin, ``Temporal Organization in Cells"
Academic Press, London (1963).

\bibitem{Goodwin-Cohen}
B. Goodwin and M. H. Cohen, J. Theor. Biol. 25 (1969) 49 
  "A Phase-shift Model for the Spatial and Temporal
   Organization of Developing Systems."



\bibitem{CF-KK}
C. Furusawa and K. Kaneko,
``Emergence of Rules in Cell Society: Differentiation, Hierarchy, and
Stability'', Bull. Math. Biol., in press

\bibitem{KK-Rel}
K. Kaneko, `` Relevance of Clustering to Biological Networks'',
Physica 75 D (1994) 55

\bibitem{KK-MF}
K. Kaneko `` Mean Field Fluctuation in Network of Chaotic Elements'',
Physica 55D (1992) 368-384

\bibitem{homeo}
K. Kaneko and T. Ikegami,  ``Homeochaos: Dynamics Stability of a
symbiotic network with population dynamics and evolving mutation rates''
Physica 56 D (1992) , 406-429

\bibitem{volvox}
D.L. Kirk and J.F. Harper,
``Genetic, Biochemical, and Molecular Approaches to Volvox Development
and Evolution", International Rev. of Cytology, 99 (1985) 217

\bibitem{Ogawa}
Ogawa, M. (1993). Review: differentiation and proliferation of 
hematopoietic stem cells. {\sl Blood} {\bf 81}, 2844


\bibitem{Akiyama}
E. Akiyama and
K.  Kaneko, ``Evolution of Cooperation, Differentiation, Complexity, and
Diversity in an Iterated Three-person Game'', Artificial Life 2 (1996)
293-304


\end{thebibliography}
\end {document}